\documentclass[12pt,preprint]{aastex}
\shortauthors{Carciofi et al. }
\shorttitle{Achernar: Rapid Polarization Variability}

\begin{document}

\title{Achernar: Rapid Polarization Variability as Evidence of Photospheric and Circumstellar Activity}

\author{A.C. Carciofi \altaffilmark{1}, 
A. M. Magalh\~aes\altaffilmark{1}, 
N. V.Leister\altaffilmark{1},
J. E. Bjorkman\altaffilmark{2},
R.S. Levenhagen\altaffilmark{3},
}   

\altaffiltext{1}{Instituto de Astronomia, Geof\'isica e Ci\^encias Atmosf\'ericas, Universidade de S\~ao Paulo, Rua do Mat\~ao 1226, Cidade Universit\'aria, S\~ao Paulo, SP 05508-900, Brazil}
\altaffiltext{2}{University of Toledo  Department of Physics \& Astronomy MS111 2801 W. 
Bancroft Street Toledo, OH 43606 USA} 
\altaffiltext{3}{Centro Universit\'ario Franciscano - UNIFRA, 
Rua dos Andradas, 1614, Santa Maria, RS  97010-032, Brazil}

\begin{abstract}
We present the results of a high accuracy ($\sigma \approx 0.005\%$) polarization monitoring of the Be Star Achernar that was carried out between July 7th and November 5th, 2006.
Our results indicate that, after a near quiescent phase from 1998 to 2002, Achernar
is presently in an active phase and has built a circumstellar disk.
We detect variations both in the polarization level and position angle in timescales as short as one hour and as long as several weeks.
Detailed modeling of the observed polarization strongly suggests that the short-term variations originate from discrete mass ejection events which produce transient inhomogeneities in the inner disk. Long-term variations, on the other hand, can be explained by the formation of an inner ring following one or several mass ejection events.
\end{abstract}


\section{Introduction \label{introduction}}

A common feature to all Be stars is the presence of ionized circumstellar gas \citep{col87}.
Despite the general consensus that this gas is mainly distributed in a circumstellar disk, 
there is still much controversy on how the disk is built.
It is generally believed that rotation plays an important role in the mass transfer between the star and the disk \citep{owo05}, but 
since evidence points to the fact that not all Be stars rotate subcritically \citep{cra05}, it follows that additional driving mechanisms must be present \citep{tow04,owo05}.

This gap in our knowledge about disk formation can be attributed, at least in part, to the lack of sufficiently strong observational constraints. When the disk is fully formed, changes in its inner part may be obscured by the large emission from the outer parts.
For this reason, the best phase to investigate the mass loss is when the disk has just started forming.



Disks of Be stars show temporal variations on several timescales. Short-term (minutes to hours) line profile variations frequently occur in early-type Be stars (e.g. \citealt{por88}).
The very short period of those variations suggests that they are associated with changes in the photospheric level or the innermost part of the disk and, thus, renders short-term variations as important properties to identify the additional mechanism(s) required for a rapidly rotating B star to become a Be star.

Several such mechanisms 
were put forward (see \citealt{por03} for an overview).
Whatever the mechanism for ejecting gas from the photosphere is, it must involve
the mass transfer from some part of the stellar photosphere to some position of the inner disk.
It follows that \emph{if we can determine observationally} the mass and geometry of the material as it is fed into the disk then we can begin to learn something about the physical mechanisms responsible for building the disk.

In this letter we provide an attempt towards this goal. We present high-precision polarimetric observations of the Be Star Achernar (HD10144, $\alpha$ Eri), with temporal resolutions of a few minutes.  
We achieved accuracies as high as  0.002\% in the linear polarization and $0.4\degr$ in position angle, depending on the night conditions. 


\section{Observations \label{observations}}

The polarimetric observations have been performed with the 1.6-m Perkin-Elmer
telescope and the 0.6-m Boller \& Chivens telescope at OPD/LNA - Brazil.
We have used a CCD camera with a polarimetric module described in \citet{m96}, consisting of a rotating half-waveplate and a calcite prism placed in the telescope beam. A typical observation consists of 8 consecutive waveplate positions (hereafter WPP) separated by 22\fdg5. 
Details of data reduction can be found in \citet{mag84}.


We define the average polarization of each WPP sequence, $\bar{P}$, as the average of the $M$ independent sets of 8 consecutive WPPs, i.e., $\bar{P} = {M}^{-1}{\sum_{j=1}^{M}P_j }$,
where $P_1$ is the polarization for WPPs 1 to 8, $P_2$ for WPPs 9 to 16, and so forth.
Similarly, $\bar{\sigma} = M^{-1} \left(\sum\sigma^2_j \right)^{1/2}$.
A summary of our polarimetric data is presented in Table~\ref{t_obs}.

In each observing run at least one polarized 
standard star was observed to calibrate the observed PA.
The instrument stability was assessed by observing targets with a polarization level similar to Achernar. They were selected from the \citet{hei00} catalog  using the following criteria: 
(i) possess a polarization of about 0.15\% known with good accuracy, and (ii) be of spectral type AV or later, to ensure that the  
polarization is purely interstellar.
The selected targets were HD 150764, HD 118507 and HD 204867.
We monitored those stars in the B filter for several hours, with a S/N better than 5000,
thus producing several independent sets of 8 consecutive WPP positions. Our results indicate that the polarization is stable within a $1\sigma$ level, even under less than ideal sky conditions. 
However, we found that the PA does vary more than $1\sigma$, but never more than $2\sigma$. From those tests we conclude that the
instrument is stable, at a minimum, to the $\approx0.003$\% level for $P$ and $\approx1.4\degr$ for the PA.

Spectroscopic observations were carried out in November  05th, 2006 with the Coud\'e spectrograph at the 1.60m telescope of OPD/LNA. We used a 600 groove/mm grating blazed at 6563 \AA~in first order, resulting in a  reciprocal dispersion of 0.24 \AA/pixel. The typical S/N ratio was 300. 

\section{Results \label{results}}

\begin{figure}[bp]
\plotone{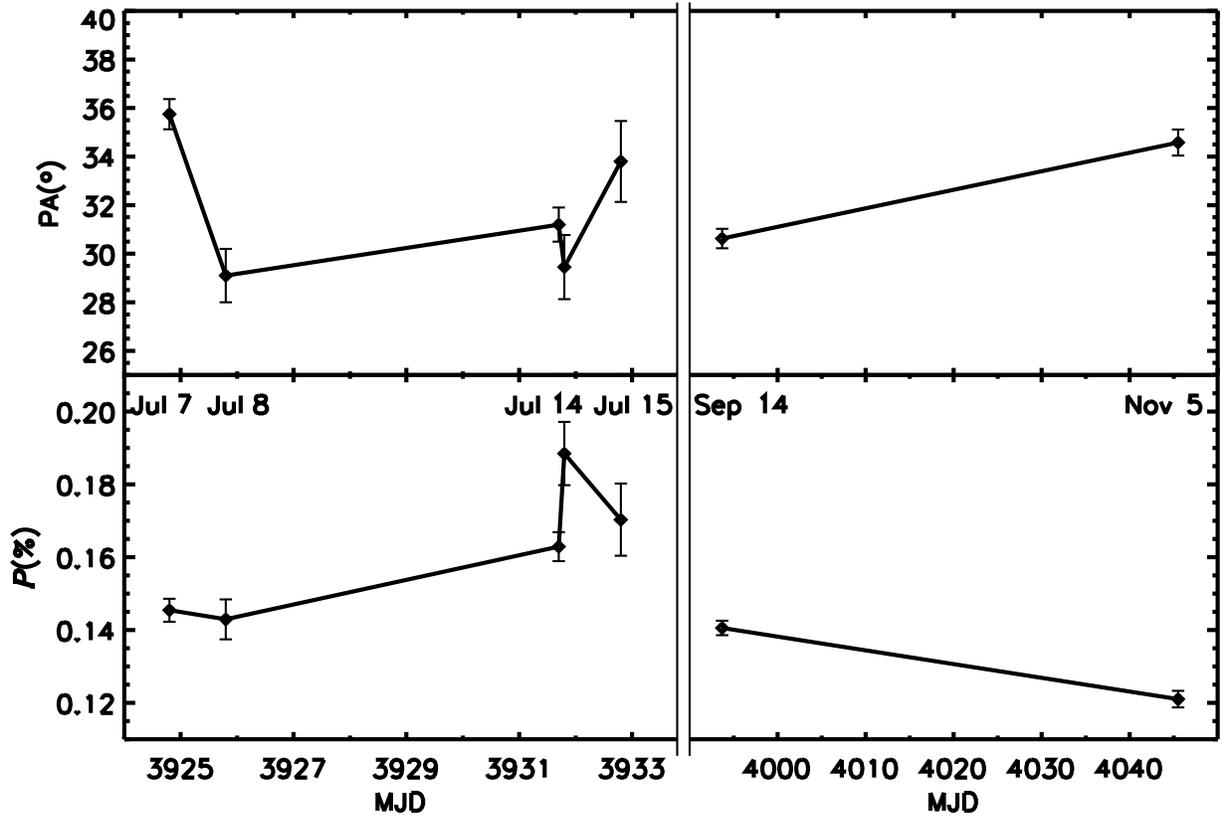}
\figcaption[]{
B-filter polarization monitoring for $\alpha$ Eri. 
\label{obs_p}}
\end{figure}

In Figure~\ref{obs_p} we show the results of our polarization monitoring in the B filter between July 7th and November 5th, 2006. 
In one week, from July 7th to 14th, $\bar{P}$ increased from $0.145 \pm 0.003$\% to $0.188\pm0.009$\%, a 30\% increase. Two months later, by September 14th, the polarization had gone back to the $0.14$\% level and by November 5th it was as low as $0.121\pm0.002$\%. From the above, we can establish that the polarization level of Achernar has varied in the period of the observations with timescales of weeks to months.

In addition to this long-term variation, the polarization seems to have varied also in much shorter timescales. 
To improve the temporal resolution of our measurements, we now group the WPPs in the following way: group 1, WPP from 1 to 8; group 2, WPP from 2 to 9; etc. This procedure effectively increases the temporal resolution, since two consecutive WPPs are separated of each other by about 4 minutes.
However, this is not the true temporal resolution, because a polarimetric measurement is obtained from 8 WPPs and, thus, encompasses about 1/2h of observations. 

\begin{figure}[bp]
\plotone{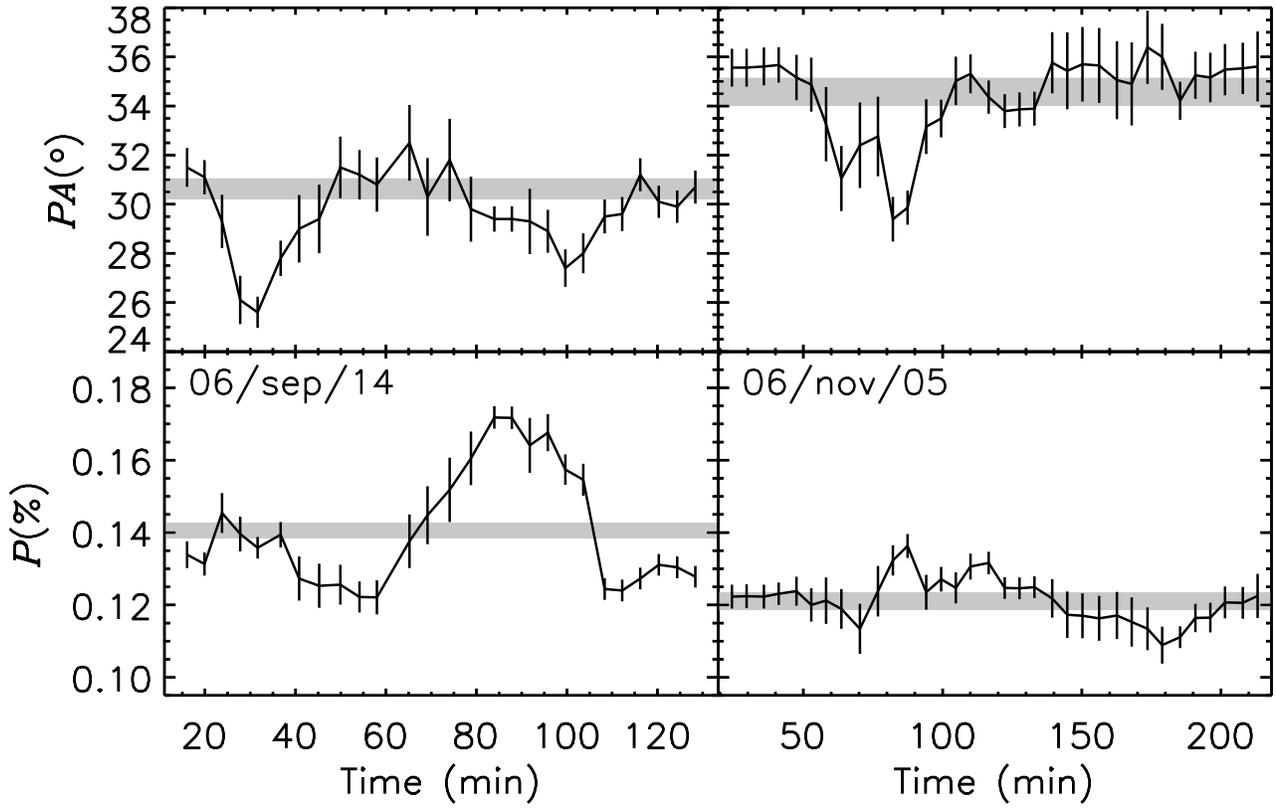}
\figcaption[]{
Short-term polarization variability. Left column: results for September 14th. Right column: results for November 5th.
The grey bands indicate the average value for the sequence, $\bar{P}$, the band thickness spanning the value of $2\bar{\sigma}$.
\label{pvar}}
\end{figure}

Our results are shown in Figure~\ref{pvar}, where we plot the polarization vs. time for the WPP sequences of September 14th and November 5th.
In September 14h, for instance, the polarization varied between $0.122\pm0.004$\% and 
$0.172\pm0.003$\% in less than one hour. 
Those values 
separated by 0.050\% which is about 12 times
larger than the typical observational uncertainty in this sequence.
There is also evidence of short-term variations in November 5th, albeit with a smaller amplitude. For the July observations there is evidence of rapid polarization variations, but the case is not so clear because of the larger observational uncertainties.


Time series analysis was performed in the polarization data using the Fast Fourier Transform with the Cleanest algorithm \citep{fos95}. Further details of the procedure can be found in \citet{vin06}.
Two frequencies with high statistical significance were found: 0.074 and 0.57 c/d. 
The later frequency is close to the frequency of 0.49 c/d found by \citet{vin06}, which they attribute to stellar rotation. This frequency, if real, is of great importance to our interpretation of the polarization variability in \S~\ref{modeling}.

In addition to variations in the polarization level, we also observed both short- and long-term variations in the PA. 
The long-term variations of the PA do not always correlate with the variations of the polarization level. For instance, from July 7th to 8th, we did not detect changes in the average polarization, but a 7\degr ($6\bar{\sigma}$) PA rotation was observed.  
In two-dimensional (2D), axisymmetric disks, the PA is expected to be constant and parallel to the symmetry axis.
The observed changes are, thus, probably associated with 
departures from a simple 2D geometry.




\section{Modeling \label{modeling}}

The above results raise the question of what are the physical mechanisms that cause the variations and whether the same mechanism is responsible for both the short- and long-term variations in the polarization.  In the following we attempt to demonstrate that the observed polarization can be explained by the presence of {\it blobs} (small regions of density enhancement) 
and/or rings of gas close  to the stellar photosphere.


Before endeavoring to quantitatively explain the variations, we must find a suitable disk model that describes reasonably well the observed polarization and H$\alpha$ line profile. For this purpose we use a recent three-dimensional non-local thermodynamic equilibrium code described in \citet{car06a} and \citet{car06b}. 
We assume that  the radial disk density is controlled by viscosity \citep{lee91} and that the disk is hydrostatic 
in the vertical direction \citep{bjo97}. Given a prescription for the gas viscosity, the disk has two free parameters: the mass loss rate, $\dot{M}$, and the outer radius, $R_d$. 

There is a wide scatter in the literature about the physical characteristics of Achernar, since its rapid rotation makes it difficult to determine its spectral type \citep{vin06}. 
We adopt an spectral type of B3V, corresponding to a $T_{\rm eff} \approx 20,000\;\rm K$ and a radius of $R_\star \approx 9 R_\sun$. We note, however, that varying those parameters within a reasonable range (say $\pm 2000\;\rm K$ for $T_{\rm eff}$ and $\pm 1R_\sun$ for $R_\star$) changes little  the results below.
$\alpha$ Eri has a strongly broadened H$\alpha$ photospheric component and this must be accounted for when fitting the observed profile. We have used the average spectrum of 1999 of \citet{vin06}, presumed to be purely photospheric, as an input photospheric profile in our modeling.

In addition to the disk and stellar parameters, one must specify the angle $i$ by which the system is observed. This parameter has been reasonably well constrained by \citet{vin06} to be around 70\degr. In our modeling we assume $i$ as a free parameter in the range 60 -- 80\degr.

\begin{figure}[bp]
\plotone{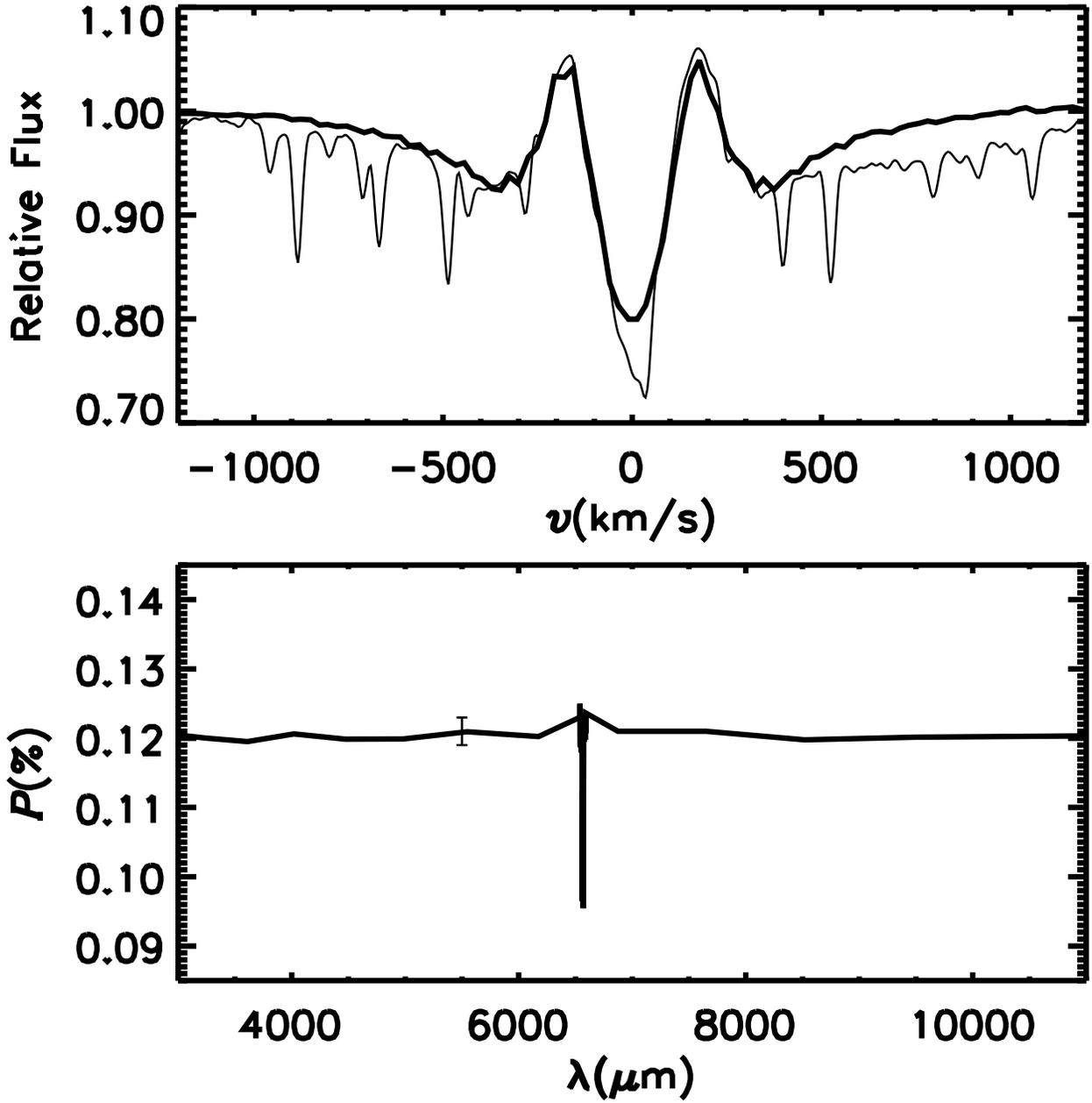}
\figcaption[]{
Best-fit model (thick lines) for the observations of November 5th. 
\label{disk_model}}
\end{figure}

We have chosen to model the observations of November 5th, since this is the only date for which we have simultaneous H$\alpha$ and polarization observations. 
Our best-fit  model 
is shown in Figure~\ref{disk_model} and the adopted model parameters are summarized in Table~\ref{t_mod}.  
We fit exactly the $B$-band polarization level and our synthetic H$\alpha$ profile reproduces well the observed profile. 

Having found a 2D model that reproduces well a subset of out observations, we now attempt to reproduce the polarization variations. Since this is an initial study, our goal is to find a qualitative explanation of the time dependence of the variations and a quantitative explanation of the amplitude of the variations.

As stated in the introduction, the driving mechanisms operating in Be stars are still unknown. For this reason we adopt an ad hoc scenario which, nevertheless, has many of the physical ingredients believed to operate in Be stars. 
We assume that after a mass mass ejection event (hereafter MEE) a region of density enhancement, which we call a \emph{blob}, is formed near the stellar photosphere. This blob is short-lived: after a few stellar periods (i.e., several days) the blob material will mix with the disk, both because of differential rotation and viscosity, thus forming a ring around the star.
The fate of the ring is controlled by viscosity: if the mass transfer from the star stops (i.e., there are no subsequent MEEs), viscosity causes inner rim of disk to lose momentum and fall back to the star and the outer rim to diffuse outwards. If other MEEs occur, the ring density and size will increase with time.

\subsection{Rings}
 
Our assumption is that the observed long-term polarization variations are due to the formation and growth of a ring after one or more mass ejection events, or to the formation of an inner hole following a period of low mass loss rate. 
To model this, we define the ring as a density enhancement in the disk between $R_\star$ and $R_r$.
The ring density is given by 
$n_r(r,z) = n_d(r,z)\gamma$, for $r< R_r$,
where $n_d$ is the disk density and $\gamma$ is the density enhancement factor. An inner hole
is represented $\gamma$ less than 1.

\begin{figure}[bp]
\plotone{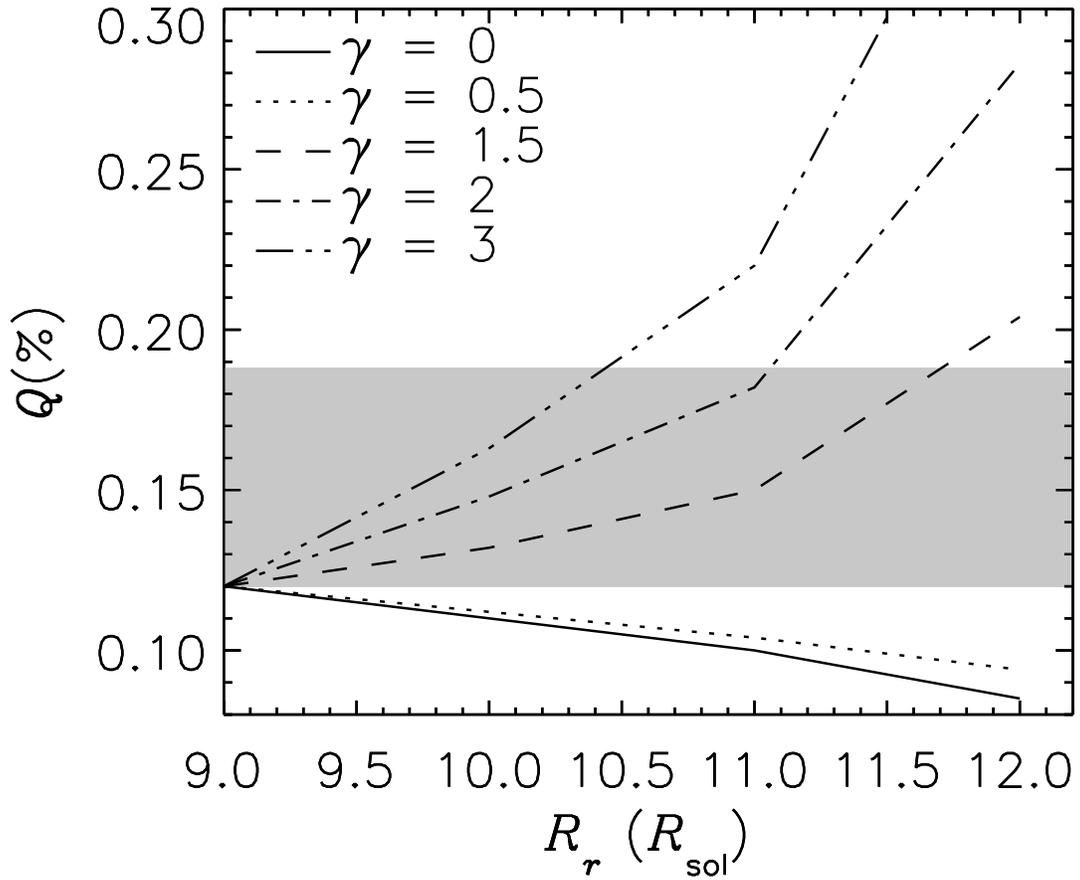}
\figcaption[]{Ring model. Shown is the $B$-band polarization vs. the ring outer radius, for several values of the density enhancement factor $\gamma$. The horizontal lines represent the minimum and maximum polarization observed by us.
\label{ring_model}}
\end{figure}

In Figure~\ref{ring_model} we show polarization results for models with $R_r$ between 10 to 12 $R_\sun$ and $\gamma$ between 0 and 3. 
From those models we learn that small rings with large density enhancements ($\gamma > 2$) successfully explain the observed amplitude of the long-term variations of the polarization. Similar effects are produced by larger rings with smaller density enhancements. 

We can estimate the time $\tau$ required for the formation of a ring of radius $R_r$ using the following expression for the viscous diffusion timescale (e.g. \citealt{bjo97})
\begin{equation}
\tau = {R_r^2}{\nu}^{-1},
\end{equation}
where $\nu$ is the kinematic viscosity of the gas. Assuming the {\it eddy viscosity} of \citet{sha73} the above equation can be written as 
\begin{equation}
\tau = {R_r^2}\left( a\alpha H\right)^{-1},
\end{equation}
where $a$ is the sound speed, $\alpha$ the viscosity parameter introduced by Shakura \& Syunyaev and $H$ is the disk scaleheight.
For pressure supported, hydrostatic and isothermal disks, $H$ can be written as
$H=aV_{\rm crit}^{-1}R_\star^{-0.5}r^{1.5}$, where $r$ is the distance to the star and $V_{\rm crit}$ is the critical rotational velocity of the star. Assuming $V_{\rm crit}\approx400 \rm km\;s^{-1}$, $a\approx20 \rm km\;s^{-1}$ and $\alpha=0.2$, we find that a $2R_\sun$-wide ring can be formed in a couple of weeks only. This means the timescale for ring formation or dissipation is of the same order of magnitude than our observed timescales for long-term polarization variations.

We should point out that the ring model cannot account for the observed changes in the PA because it is intrinsically axisymmetric. In addition, it cannot explain the observed short-term variations, since the associated timescale of ring formation/dissipation (weeks) is much larger than the typical timescale of the short-term variations (hours).


\subsection{Blobs}


We now attempt to quantify the polarization effects of an asymmetric gas distribution in the inner disk assuming a semi-spherical blob of material attached to the stellar photosphere at the equator. For simplicity we assume the blob has a constant density $n_\mathrm{blob}$.

\begin{figure}[bp]
\plotone{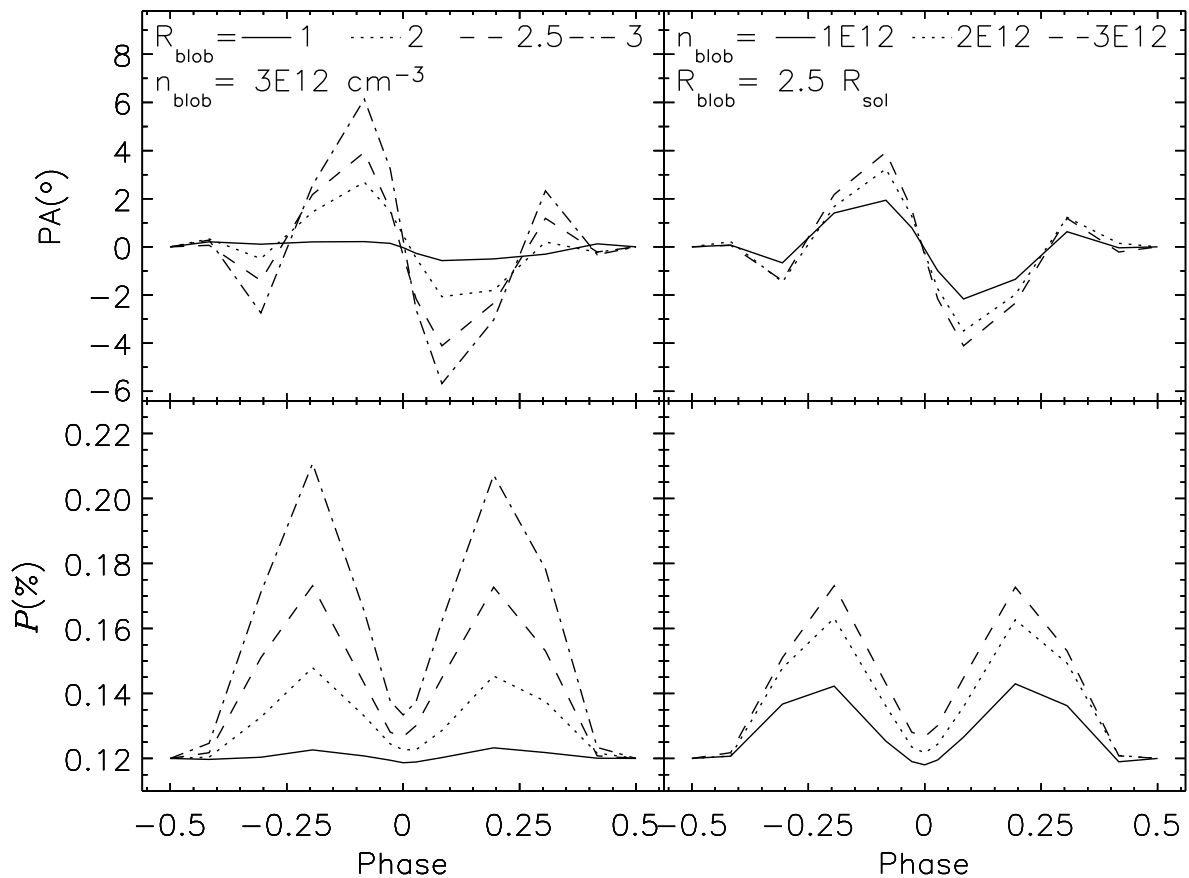}
\figcaption[]{Blob model. $B$-band polarization (bottom) and position angle (top)  vs. orbital phase.
The models on the left show the effects of blob radius ($R_{\rm blob} = 1$ to $3\;R_\sun$) for a given blob density ($n_{\rm blob} = 3\times10^{12} \rm cm^{-3}$).
The models on the right show the effects of blob density ($n_{\rm blob}$ between 1 and $3\times10^{12} \rm cm^{-3}$) for a given blob radius ($R_{\rm blob} = 2.5\;R_\sun$).
A phase of zero corresponds to the blob between the star and the observer.
\label{blob_model}}
\end{figure}

In Figure~\ref{blob_model} we show the polarization level and PA as a function of the blob orbital phase  for several blob sizes and densities. 
We adopt blob radii between 1 and $3R_\sun$ and $n_\mathrm{blob}$ between 1 and $3 \times 10^{12}\;\rm cm^{-3}$, which corresponds to 1.5 -- 4.5 times the density at the base of the disk, $n_0$. 

Our results show that blobs with the above characteristics can successfully explain the amplitude of observed short-term variations in the polarization level. For instance, the $\Delta P\approx 0.05\%$ variation observed in September 14th can be accounted for by most models of Figure~\ref{blob_model}.
More importantly, however, the blob model can also reproduce the amplitude of the observed changes in PA.

The blob model has two distinct associated timescales. The first is, clearly, the rotational period of the star, since in our assumption the blob is initially attached to the stellar photosphere.
The orbital period of Achernar is very short (about 2 days, \citealt{vin06}), and from Figure~\ref{blob_model} we obtain that  $6 \degr$ changes in the PA or $0.05\%$ changes in the polarization level can occur in $\Delta\phi<0.1$ or about 4 hours.

The second timescale is associated with the process of the blob formation and growth during a MEE.
It is difficult to estimate a timescale for this situation since we do not know how the blob forms;
however, if we arbitrarily, for the sake of argument, give an initial velocity of a few hundred km/s to the blob material, we find that the blob can become as large as $1R_\sun$ in less than one hour.

The above discussion is very qualitative and must be verified by detailed calcutations.
The important point, however, is that the blob model appears to have associated timescales of one to a few hours, in accordance to our observations.

The structures of the proposed scenario have a clear hierarchy: MEEs give rise to blobs which form rings which, in turn, form the disk. We cannot yet, with the present data, determine important quantities such as the characteristics (mass, geometry and rate) of the MEEs. 
However, from our modeling the typical masses of the structures involved can be roughly estimated.
Our best-fit 2D disk model has a mass of $3\times10^{-10}\;M_\sun$. 
A ring of radius $R_r=11\;R_\sun$ and $\gamma=2$, which can produce a polarization variation in the observed range, corresponds to a mass input to the disk of about 7\% of the total disk mass.
Similarly, a representative blob model, say with $R_\mathrm{blob} = 2\;R_{\sun}$ and $n_\mathrm{blob} = 3 \times 10^{12}\;\rm cm^{-3}$, corresponds to a mass input to the disk of about 5\% of the disk mass.

\section{Conclusions \label{Conclusions}}

We report the detection of both short- and long-term variations in the polarization level and position angle of Achernar. 
From an initial study using a state-of-the-art computer code we arrived at a very plausible picture to explain the observed variations.
In the proposed scenario, short-term variations are linked to two processes: the formation of a blob following a mass ejection event and the co-rotation of this blob with the (supposedly Keplerian) velocities of the inner disk. The finding of a significant frequency of 0.57 c/d in the polarization data, very close to the stellar rotation period, 0.49 c/d, strongly supports our assumption that the short-period variations are linked to photospheric processes.


The long-term variations are probably a combinations of two processes: the formation of blobs and their subsequent dissipation to form a ring around the star.
The ring scenario has already been proposed by \citet{riv01} to explain line profiles variations observed in several early-type Be stars. Since this scenario is a natural consequence of the viscous decretion model combined with a non-constant mass loss rate, we suggest that observations aiming at detecting and characterizing rings in the inner disk of Be stars will constitute an important test of the viscous decretion model.


The determination of the rate and typical masses of the mass ejection events would be a very important observational result. This cannot be achieved with our present data, but our results above strongly suggest that this determination should be possible with a long-term, high-precision polarization and spectroscopic monitoring of the early phase of the disk formation.

\clearpage

\begin{deluxetable}{lccc}
\tablecaption{Polarimetric Observations. \label{t_obs}}
\tablewidth{0pt}
\tablehead{
\colhead{Date} &
\colhead{Nr. of} &
\colhead{$\bar{P}$} &
\colhead{$\bar{\mathrm{PA}}$} 
\\
\colhead{(2006)} &
\colhead{WPPs} &
\colhead{(\%)} &
\colhead{(\degr)}
}
\startdata
Jul. 07 & 
32 
& $ 0.145 \pm 0.003 $ & $ 35.8 \pm 0.6 $ \\

Jul. 08 & 
24 
& $ 0.143 \pm 0.005 $ & $ 29.1 \pm 1.1 $    \\

Jul. 14 & 
8 
& $ 0.163 \pm 0.004 $ & $ 31.2 \pm 0.4 $   \\

Jul. 14 & 
20 
& $ 0.188 \pm 0.009 $ & $ 29.5 \pm 1.3 $   \\



Jul. 15 & 
16 
& $ 0.17 \pm 0.01 $ & $ 34 \pm 2 $ \\


Sep. 14 & 
34 
& $ 0.141 \pm 0.002 $ & $ 30.6 \pm 0.4 $ \\

Nov. 05 & 
41 
& $ 0.121 \pm 0.002 $ & $ 34.6 \pm 0.5 $ \\
\enddata
\end{deluxetable}


\clearpage
\begin{deluxetable}{ccccccc}
\tablecaption{Adopted Stellar Parameters and Fitted Disk Parameters for Achernar \label{t_mod}}
\tablewidth{0pt}
\tablehead{
\colhead{$R_{\star}$} &
\colhead{$T_\mathrm{eff}$} &
\colhead{$V_\mathrm{crit}$} &
\colhead{$\dot{M}$} &
\colhead{$n_0$} &
\colhead{$R_d$} &
\colhead{$i$}
\\
\colhead{$(R_{\sun})$} & 
\colhead{(K)} & 
\colhead{(km\,s$^{-1}$)} & 
\colhead{$(M_{\sun}\;\rm yr^{-1})$} &
\colhead{$(\rm cm^{-3})$} &
\colhead{$(R_{\star})$} & 
\colhead{(\degr)}
}
\startdata
9 & $20,000$ & 400 & $8\times 10^{-12}$ & $6.7\times10^{-11}$ &15 & 65 \\
\enddata
\end{deluxetable}

\clearpage


\begin{thebibliography}{}


\bibitem[Bjorkman(1997)]{bjo97} Bjorkman, J.~E.\ 1997, Circumstellar Disks, in Stellar Atmospheres: Theory and Observations, ed. J.~P.~de~Greve, R.~Blomme, 
\& H.~Hensberge (New York: Springer)


\bibitem[Carciofi \& Bjorkman(2006)]{car06a} Carciofi, A.~C., 
\& Bjorkman, J.~E.\ 2006, \apj, 639, 1081 

\bibitem[Carciofi et al.(2006)]{car06b} Carciofi, A.~C., et 
al.\ 2006, \apj, 652, 1617 

\bibitem[Collins (1987)]{col87} Collins, G. W. 1987, in IAU Colloq. 92, Physics of Be Stars, ed. 
A. Slettebak \& T. P. Snow (Cambridge: Cambridge Univ. Press), 3 


\bibitem[Cranmer(2005)]{cra05} Cranmer, S.~R.\ 2005, \apj, 634, 585 

\bibitem[Foster (1995)]{fos95} Forster, G. 1995, \aj, 109, 1889

\bibitem[Heiles(2000)]{hei00} Heiles, C.\ 2000, \aj, 119, 923 


\bibitem[Lee et al.(1991)]{lee91} Lee, U., Osaki, Y., \& 
Saio, H.\ 1991, \mnras, 250, 432 

\bibitem[Magalh\~aes et al.(1984)]{mag84} Magalh\~aes, A.~M., 
Benedetti, E., \& Roland, E.~H.\ 1984, \pasp, 96, 383 

\bibitem[Magalh\~aes et al. (1996)]{m96} Magalh\~aes A.~M., Rodrigues C.~V., Margoniner V.~E.,
Pereyra A., Heathcote S., 1996, in  ASP Conf. Ser. 97, Polarimetry of the Interstellar Medium, ed. Roberge W.~G. \& Whittet D.~C.~B. (San Francisco: ASP), 118



\bibitem[Porter \& Rivinius(2003)]{por03} Porter, J.~M., \& Rivinius, Th.
\ 2003, PASP, 115, 1153

\bibitem[Owocki(2005)]{owo05} Owocki, S.\ 2005, in ASP Conf. Ser. 337, The Nature 
and Evolution of Disks Around Hot Stars, ed. R. Ignace and K. G. Gayley
(San Francisco: ASP), 101

\bibitem[Porri \& Stalio(1988)]{por88} Porri, A., \& Stalio, 
R.\ 1988, \aaps, 75, 371 

\bibitem[Rivinius et al.(2001)]{riv01} Rivinius, T., Baade, 
D., {\v S}tefl, S., \& Maintz, M.\ 2001, \aap, 379, 257



\bibitem[Shakura \& Syunyaev(1973)]{sha73} Shakura, N.~I., \& 
Syunyaev, R.~A.\ 1973, \aap, 24, 337 

\bibitem[Townsend et al.(2004)]{tow04} Townsend, R.~H.~D., 
Owocki, S.~P., \& Howarth, I.~D.\ 2004, \mnras, 350, 189 

\bibitem[Vinicius et al.(2006)]{vin06} Vinicius, M.~M.~F., 
Zorec, J., Leister, N.~V., \& Levenhagen, R.~S.\ 2006, \aap, 446, 643 



\end{thebibliography}
\end{document}